\documentclass[superscriptaddress,pra, amsmath,amssymb,reprint,showkeys,showpacs]{revtex4-1} %%%Dos Columnas

\usepackage{graphicx}
\usepackage{graphics}
\usepackage{amssymb}
\usepackage{amsmath,bm}
\usepackage{float}
\usepackage[usenames,dvipsnames]{color}
\usepackage[export]{adjustbox}
\usepackage{hyperref}
\hypersetup{
    colorlinks=true,
    linkcolor=blue,
    citecolor=blue,
    filecolor=cyan,      
    urlcolor=cyan,
}
\begin{document}

\title{Alignment-dependent decay rate of an atomic dipole near an optical nanofiber}

\author{P. Solano}
 \email{Corresponding author email: solano.pablo.a@gmail.com}
\affiliation{Joint Quantum Institute and Department of Physics, University of Maryland, College Park, MD 20742, USA.}

\author{J. A. Grover}
\affiliation{Joint Quantum Institute and Department of Physics, University of Maryland, College Park, MD 20742, USA.}

\author{Y. Xu}
\affiliation{Department of Electrical and Computer Engineering and 
the Institute for Research in Electronics and Applied 
Physics, University of Maryland, 
College Park, Maryland 20742-3511, USA.}

\author{P. Barberis-Blostein}
\affiliation{Joint Quantum Institute and Department of Physics, University of Maryland, College Park, MD 20742, USA.}
\affiliation{Instituto de Investigaciones en Matem\'{a}ticas Aplicadas y en Sistemas, Universidad Nacional Aut\'{o}noma de M\'{e}xico, Ciudad Universitaria, 04510, DF, M\'{e}xico.}

\author{J. N. Munday}
\affiliation{Department of Electrical and Computer Engineering and 
the Institute for Research in Electronics and Applied 
Physics, University of Maryland, 
College Park, Maryland 20742-3511, USA.}

\author{L. A. Orozco}
\affiliation{Joint Quantum Institute and Department of Physics,
University of Maryland, College Park, MD 20742, USA.}

\author{W. D. Phillips}
\affiliation{Joint Quantum Institute, NIST and University of Maryland, Gaithersburg, Maryland 20899, USA}

\author{S. L. Rolston}
\affiliation{Joint Quantum Institute and Department of Physics,
University of Maryland, College Park, MD 20742, USA.}

\date{\today}

\begin{abstract}
We study the modification of the atomic spontaneous emission rate, i.e. Purcell effect, of $^{87}$Rb in the vicinity of an optical nanofiber ($\sim$500 nm diameter). We observe enhancement and inhibition of the atomic decay rate depending on the alignment of the induced atomic dipole relative to the nanofiber. Finite-difference time-domain simulations are in quantitative agreement with the measurements when considering the atoms as simple oscillating linear dipoles. This is surprising since the multi-level nature of the atoms should produce a different radiation pattern, predicting smaller modification of the lifetime than the measured ones. This work is a step towards characterizing and controlling atomic properties near optical waveguides, fundamental tools for the development of quantum photonics. 
 \end{abstract}

\maketitle

\section{Introduction}

Neutral atoms coupled to optical waveguides is a growing field of research \cite{thompson_coupling_2013, petersen_chiral_2014, mitsch_quantum_2014, goban_superradiance_2015, goban_atomlight_2014, yalla_efficient_2012, junge_strong_2013, volz_nonlinear_2014, patel_efficient_2016, rosenblum_extraction_2016}. Atom-waveguide systems enable atom-light interaction for propagating light modes. This makes them promising tools for forthcoming optical technologies in the quantum regime, such as quantum switches \cite{oshea_fiber-optical_2013, shomroni_all-optical_2014, tiecke_nanophotonic_2014}, diodes \cite{sayrin_nanophotonic_2015, shen_single-photon_2011}, transistors \cite{chang_single-photon_2007}, and electromagnetically induced transparency and quantum memories \cite{gouraud_demonstration_2015, sayrin_storage_2015, jones_ladder-type_2015, kumar_multi-level_2015}. In order to further any of these applications it is necessary to understand and control the effects of such waveguides on nearby atoms. 

Two important features result from having a waveguide with a preferential optical mode: the spatial variation of the electromagnetic field, and the change of its density of modes per unit frequency. One of the key atomic properties affected by both is the spontaneous emission rate \cite{fermi}. Its modification is due to the change in the local vacuum field felt by the atom under the boundary condition imposed by the adjacent object, a phenomena known as Purcell effect \cite{PhysRev.69.674.2}. When the symmetry of the free-space vacuum field is broken in the presence of an object, the alignment of the atomic dipole relative to the object also plays an important role on the atomic lifetime. For a given alignment the atom can couple more strongly (weakly) to the vacuum modes, producing an increase (decrease) of the spontaneous emission rate. The effect of waveguides on the spontaneous emission of nearby emitters has been a productive field of research \cite{le_kien_spontaneous_2016, bordo_purcell_2012, barthes_purcell_2011, lodahl_controlling_2004, PhysRevLett.95.013904, RevModPhys.87.347, le_feber_nanophotonic_2015, Nguyen2008, Nha1998, :/content/aip/journal/apl/94/10/10.1063/1.3098072, Eggleston10022015, PhysRevLett.58.2059,ropp2015}. 

Optical nanofiber (ONF) waveguides \cite{morrissey_spectroscopy_2013, chen_review_2013} are optical fibers with a diameter smaller than the wavelength of the guided field. Most of the electromagnetic field propagates outside the dielectric body of the ONF (in vacuum) in the form of an evanescent field, and its strong transversal confinement enables interactions with adjacent atoms. The nanofiber is adiabatically connected, through a tapered section, to a conventional single mode optical fiber, facilitating light coupling and readout. ONFs are up-coming platforms for photonic based quantum technologies due to the coupling efficiency of light, high surface quality at the nanometer scale, and simplicity and robustness of the fabrication procedure \cite{hoffman_ultrahigh_2014}. The electromagnetic mode confinement in an ONF provides a large atom-light coupling \cite{kien_field_2004, qi_dispersive_2016}, a feature that has been used for spectroscopy \cite{Nayak20124698}, atomic cloud characterizations \cite{PhysRevA.92.013850, morrissey_tapered_2009} and atom trapping \cite{vetsch_optical_2010,reitz_coherence_2013, kato_strong_2015, goban_demonstration_2012, PhysRevLett.113.263603, PS2017}. It also allows the operation and control of memories \cite{gouraud_demonstration_2015, sayrin_storage_2015, jones_ladder-type_2015, kumar_multi-level_2015}, and light reflectors \cite{corzo_large_2016, sorensen_coherent_2016} at the level of single photons. The presence of three polarization components of the propagating field gives rise to chiral effects and new possibilities for atom-light directional coupling including optical isolators \cite{petersen_chiral_2014, mitsch_quantum_2014, sayrin_nanophotonic_2015, lodahl_chiral_2016}. 

The Purcell effect experienced by an emitter near an ONF has been studied in the past \cite{klimov_spontaneous_2004, le_kien_spontaneous_2005, le_kien_anisotropy_2014, le_kien_scattering_2006, kien_cooperative_2008, le_kien_effect_2008, hakuta_manipulating_2012, almokhtar_numerical_2014, verhart_single_2014, doi:10.1021/acsnano.6b02057}.  
However, there are disagreements between predicted values for the decay rates (\textit{e.g.} Refs. \cite{le_kien_spontaneous_2005} and \cite{verhart_single_2014} differ by approximately 30\% for atoms at the ONF surface), without direct experimental evidence that allows to validate one calculation over the other. Moreover, the possibility of controlling the atomic lifetime in the vicinity of an ONF by the position and alignment of the emitter has not been emphasized or shown experimentally.

We measure the modification of the spontaneous emission decay rate of a $^{87}$Rb atom placed near an ONF, in the time domain for different alignments of the induced atomic dipole, showing that the atomic lifetime can increase or decrease by properly preparing the atom. We present a theoretical description of the system, and perform both finite-differences time-domain (FDTD) and electromagnetic modes expansion calculations of the modification of the atomic decay rate. The FDTD numerical calculations considering a simple two-level atom show quantitative agreement with our experimental result. However, given the the multi-level structure of the atoms, their radiation patterns should differs from that of a linear dipole. The more isotropic pattern of our multilevel atom raises a puzzling question about the interpretation of the measured effects.  Nonetheless, this study offers insight about the possibility of controlling atomic properties near surfaces for photonics, quantum optics and quantum information applications.

This paper is organized as follows: Sec. \ref{description} explains the platform under study. The details of the experimental apparatus and the measurements procedure are in Sec. \ref{apparatus}, and the results are discussed in Sec. \ref{data}. We present numerical calculations for the atomic decay rate under the experimental conditions in Sec. \ref{NS} and a theoretical modeling of the system in Sec. \ref{TM}. We comment on the role of the multilevel structure of real atoms in our experiment in Sec. \ref{multilevel}. Sec. \ref{comparison} presents a quantitative comparison of the results to numerical simulations. Finally, we discuss the implications of this result in Sec. \ref{discussion}, and conclude in Sec. \ref{conclusion}.
%%%%%%%%%%%%%%%%%%%%%%%%%%%%%%%%%%%%%%%%%%%%%%%%%%%%%

\section{Description of the experiment }\label{description}

\begin{figure}
\includegraphics[width=0.45\textwidth]{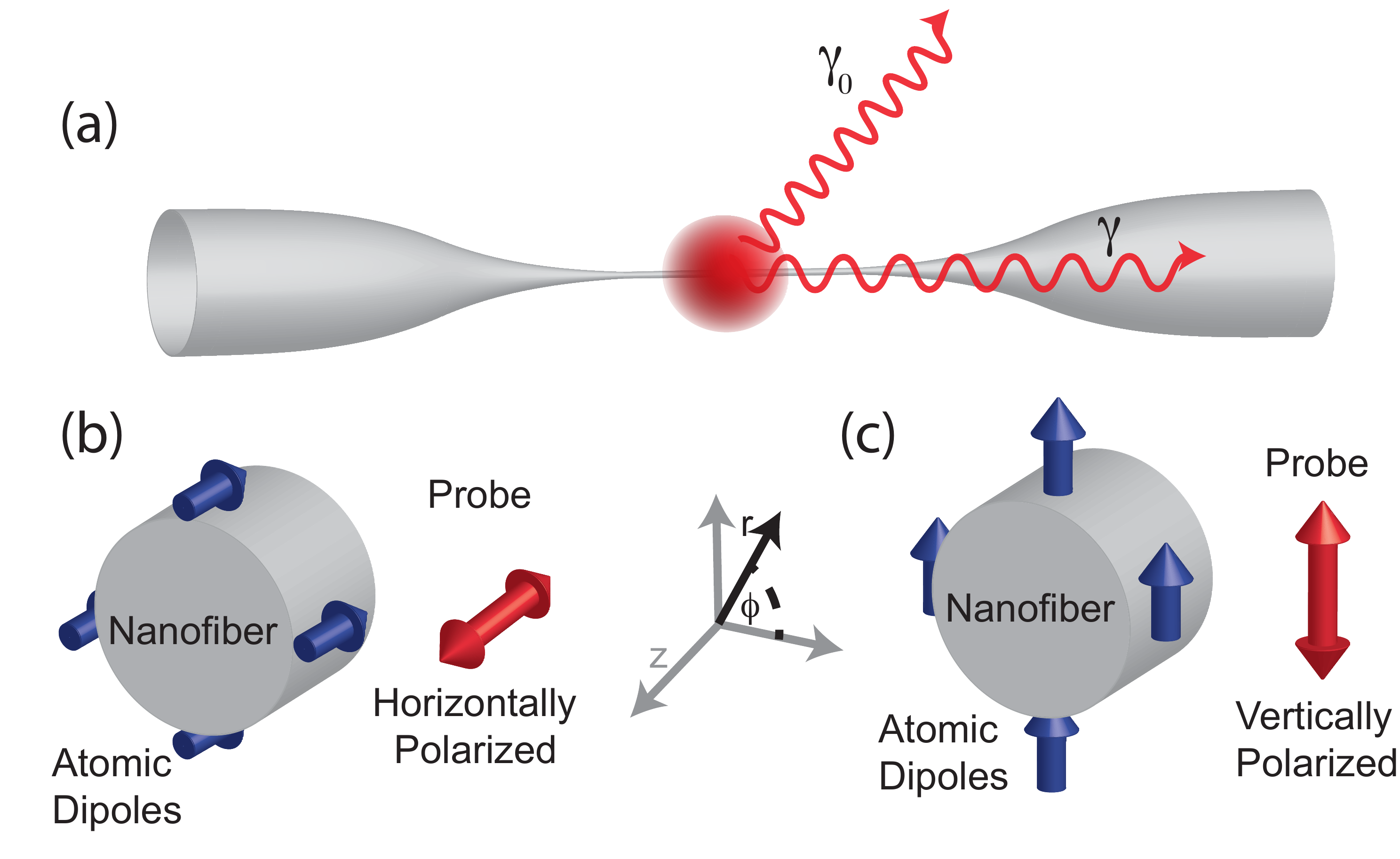}
\caption{(a) Sketch of the experimental configuration where an ensemble of cold atoms spontaneously emit photons at a rate $\gamma_{0}$ or $\gamma$ when they are placed far away or close the the fiber respectively. (b) and (c) sketch of the orientation of the induced atomic dipoles relative to the nanofiber for horizontal and vertical probe beam polarization respectively. The coordinate system is used throughout the paper.}
\label{sketch}
\end{figure}

We consider an ONF that only allows the propagation of the fundamental mode $\mathit{HE}_{11}$.  Excited atoms that are close to the nanofiber can spontaneously emit not only into free space, but also into the ONF mode, as sketched in Fig. \ref{sketch} (a). Our goal is to measure the modified spontaneous emission rate $\gamma$ of an atom placed near it, compared to the free space decay rate  $\gamma_{0}$.  $\gamma$ is the sum of the spontaneous emission rate of photons radiated into free-space (in the presence of the ONF) and into the ONF waveguide, i.e. $\gamma(\mathrm{\bold{r}})=\gamma_{\text{fs}}(\mathrm{\bold{r}})+\gamma_{\text{wg}}(\mathrm{\bold{r}})$, where all the quantities are a function of the atom position $\mathrm{\bold{r}}$. When the atom is placed far away from the ONF $\gamma_{\text{wg}}\rightarrow0$ and $\gamma\rightarrow\gamma_0$, recovering the free space scenario.

The atomic decay rate can be calculated from Fermi's golden rule \cite{fermi}. It states that the decay rate from an initial state $|i\rangle$ to a final state $|f\rangle$ is given by the strength of the interaction that mediates the transition, related to $H_{\text{int}},(\mathrm{\bold{r}})$ and the density of final states per unit energy $\rho(\epsilon)$ as
\begin{equation}
\gamma_{i\rightarrow f} (\mathrm{\bold{r}}) = \frac{2\pi}{\hbar} \rho(\epsilon) |\langle f| H_{\text{int}}(\mathrm{\bold{r}})|i\rangle|^2,
\label{FGR}
\end{equation}
where $\hbar$ is the reduced Plack constant. In our particular case $H_{\text{int}}(\mathrm{\bold{r}}) = \bm{d} \cdot \bm{E}(\mathrm{\bold{r}})$, given by the transition dipole moment $\bm{d}$ and the electric field operator $\bm{E}(\mathrm{\bold{r}})$.

The effect of the ONF dielectric body on the decay rate of a nearby atom can be thought in two analogous ways \cite{hinds_cavity_1990}: it modifies the structure of the vacuum electric field; or it reflects the emitted field back to the atom. In both cases the electric field $\bm{E}(\mathrm{\bold{r}})$  at the position of the atom is modified. This changes the interaction Hamiltonian $H_{\text{int}}(\mathrm{\bold{r}})$ along with the decay rate. The dot product between the atomic dipole moment and the electric field in Eq. (\ref{FGR}) depends upon their relative alignments, leading to alignment dependence of the atomic decay rate, because the ONF breaks the isotropy of the free space field. 

A linearly polarized optical field will drive a two-level atom along the direction of light polarization. After a scattering event, the light will leave the atom with the polarization and radiation pattern of a classical dipole aligned in such direction. By choosing the direction of light polarization we can align the radiating dipole relative to the ONF  (see Fig. \ref{sketch} (b) and (c)). This allows us to observe the dependence of the atomic decay rate on the dipole orientation. Due to the tight transverse confinement of the light propagating through the ONF, the electric field has a significant vector component along the propagation axis, as well as perpendicular to it \cite{le_kien_spontaneous_2005}. This enables an atomic dipole oscillating along the ONF to couple light into the guided mode. This is not the case for radiation in free space, where there is no radiated power along the dipole axis \cite{corney}.

Modifications in the spontaneous emission rate change the atomic spectral width and can be measured in frequency space by doing precision spectroscopy \cite{PhysRevLett.76.2866}. However, the atomic spectrum is highly susceptible to broadening mechanisms such as Stark, Zeeman (DC and AC) and Doppler shifts, and van der Waals effects from the ONF dielectric surface. These broadenings increase systematic errors, making the measurement more challenging. Considering this, we perform a direct atomic lifetime measurement, \textit{i.e.} in the time domain, to study the atomic decay rates. 

Atoms have to be relatively close to the ONF surface (less than $\lambda/2\pi$) when we probe them to see a significant effect. Two-color dipole traps, created by the evanescent field of an ONF, are a useful tool for trapping a large number of atoms close to the nanofiber \cite{reitz_coherence_2013, kato_strong_2015, goban_demonstration_2012, PhysRevLett.113.263603, corzo_large_2016, sorensen_coherent_2016, PS2017}. However, the created potential minimum is usually too far from the ONF surface (typically $\sim 200$ nm) to observe changes in the atomic radiative lifetime. Cold atoms that are free to move can get much closer to the ONF and spend sufficient time around it to be properly measured. 

To measure $\gamma/\gamma_{0}$ we overlap a cold cloud of atoms with a single mode ONF (see Fig.\ref{sketch} (a)). The atoms in the cloud are excited by a resonant probe pulse propagating perpendicularly to the nanofiber. After the pulse is suddenly turned off, spontaneously emitted light is collected and the photon-triggered signals are counted and histogrammed to get their temporal distribution, a technique known as time-correlated single photon counting (TCSPC) \cite{TCSPC}. From the exponential decay of the temporal distribution of photons we measure the atomic lifetime $\tau=1/\gamma$, directly related to the spontaneous emission rate. By detecting the spontaneously emitted light coupled into the ONF mode we are measuring only those atoms that are close enough to the nanofiber to couple light in. This allows us to obtain the modified spontaneous emission rate of atoms near the ONF surface. Note that the measured decay is the total decay rate $\gamma$, regardless of the mode used for the detection. The decay rates into different channels, in our case $\gamma_{\text{fs}}$ and $\gamma_{\text{wg}}$, only determine the branching ratio of the total decay.

We are interested in the effect of the atomic dipole alignment relative to the ONF. For this we externally drive the atomic dipole in a particular direction set by the polarization of the probe pulse. That polarization can be set to be linear in the direction along the ONF (horizontally polarized) or perpendicular to it (vertically polarized). When probing with horizontally polarized light the atomic dipoles for two-level atoms are oriented along $z$, (see  Fig. \ref{sketch} (b)). For the case of a vertically polarized probe, the atomic dipoles are oriented along $r$ on top and bottom, but along $\phi$ on each side, relative to the direction of propagation of the probe. In the vertical polarization case, we have a continuous distribution of dipole alignments, from dipoles along $r$ to dipoles along $\phi$ (see  Fig. \ref{sketch} (c)).

%%%%%%%%%%%%%%%%%%%%%%%%%%%%%%%%%%%%%%%%%%%%%%%%%%%%%
\section{Apparatus and Measurements Procedure} \label{apparatus}

\begin{figure}
\includegraphics[width=0.45\textwidth]{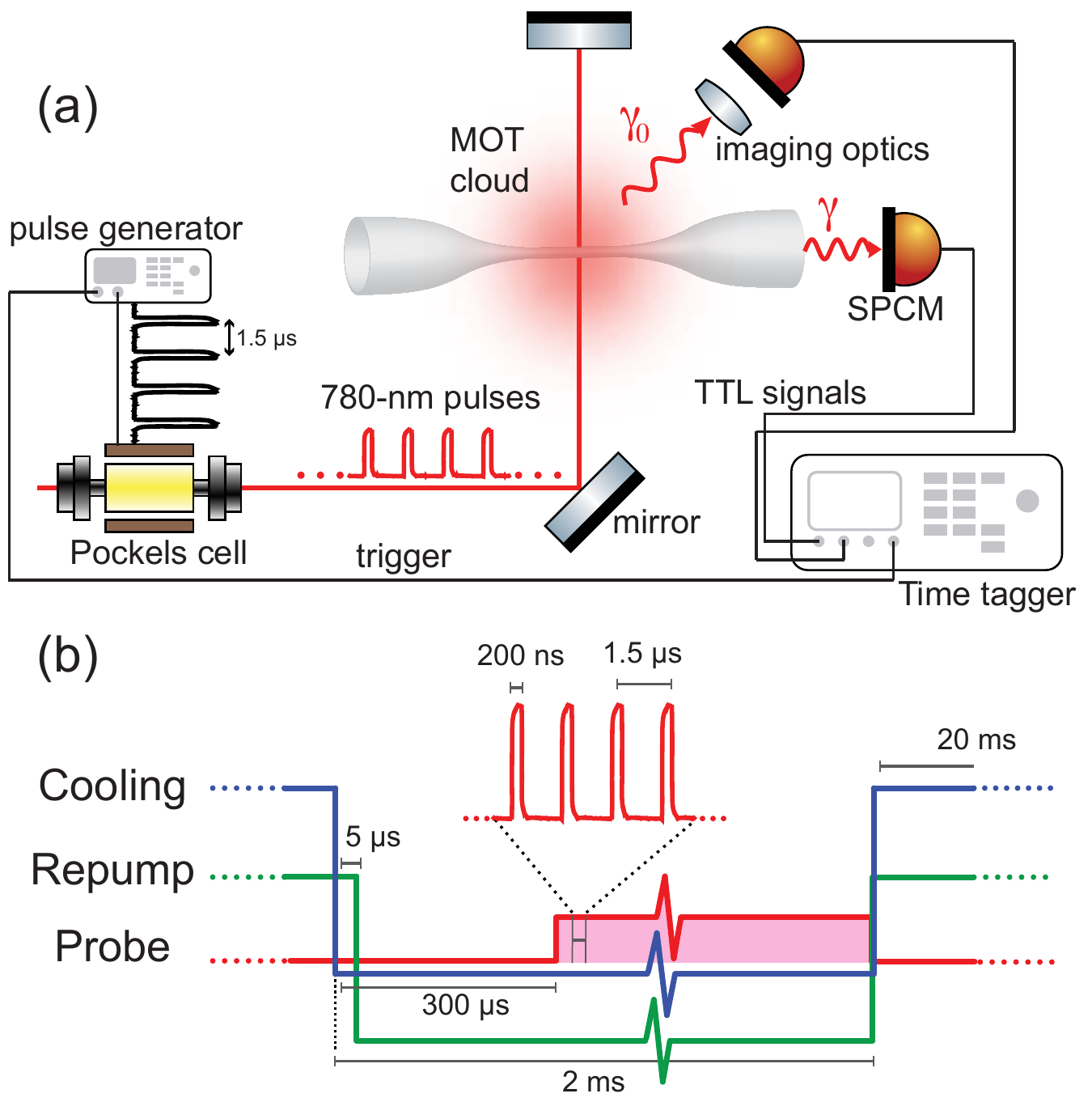}
\caption{(a) Schematic of the experimental setup. A train of pulses generated from a Pockels cell is directed to an ensemble of cold $ ^{87}$Rb atoms placed near an optical nanofiber. The spontaneously emitted photons into the nanofiber are collected and time tagged to obtain the atomic radiative lifetime. Photons emitted into free space are also measured to verify possible systematic errors. (b) Experimental sequence of light pulses to cool, repump and probe the atoms. }
\label{setup}
\end{figure}

Figure \ref{setup} (a) shows a schematic of the experimental apparatus. The ONF waist is $7$ mm in length with an approximately $240 \pm 20$ nm radius, where the uncertainty represents the variation in any given fabrication of an ONF, as destructively measured by a scanning electron microscope and independently confirmed with non-destructive techniques \cite{Fatemi:17}. Any given ONF is uniform to within 1\% through its full length. We placed the ONF inside an ultrahigh vacuum (UHV) chamber. Inside the chamber, the ONF is overlapped with a cloud of cold $ ^{87}$Rb atoms created from a magneto optical trap (MOT), loaded from a background gas of atoms released from a dispenser. The atoms are excited by pulses of a probe beam incident perpendicularly to the nanofiber and retroreflected to reduce photon-to-atom momentum transfer. These pulses are resonant with the $F=2 \rightarrow F'=3$ transition of the D2 line and created with a Pockels cell (Conoptics 250-160) for a fast turn off,  with a pulse extinction ratio of 1:170 in $20$ ns. The on-off stage of the pulses is controlled with an electronic pulse generator (Stanford Research Systems DG645). The probe beam is a 7 mm $1/e^2$ full-width collimated beam and kept at a saturation parameter $s<0.05$ to reduce atomic excitations during the off period (where $s=I/I_{\text{sat}}=2(\Omega/\gamma_0)^2$, with $I_{\text{sat}}=3.58$  mW cm$^{-2}$ the average saturation intensity for a uniform sub-level population distribution over all $m_F$ in $F=2$, and $\Omega$ is the on-resonance excitation Rabi frequency). A linear polarizer with extinction ratio of $10^{5}:1$ sets the probe polarization for driving the atomic dipoles along a particular direction. Any atoms in the cloud, close or far from the ONF, can be excited. The photons emitted into the nanofiber and those emitted into free space are independently collected with avalanche photodiodes (APDs, Laser Components COUNT-250C-FC, with less than 250 dark counts per second). The TTL pulses created from photons detected by the APDs are processed with a PC time-stamp card (Becker and Hickl DPC-230) and time stamped relative to a trigger signal coming from the pulse generator. We detect of the order of $10^{-3}$ photons per probe pulse, consistent with considering atomic excitation probability, coupling into the ONF, power losses through band-pass filters and other optical elements, and detection efficiencies.

The experimental cycle is described in Fig. \ref{setup} (b). Acousto-optic modulators (AOMs) control the amplitude and frequencies of the MOT and repump beams. After the atomic cloud reaches steady state, the MOT cooling and repump beams are turned off with a fall time of less than 0.5 $\mu$s. The repump turns off 5 $\mu$s after the cooling beams to end with the maximum number of atoms in the $F=2$ ground state. We wait 300 $\mu$s until the AOMs reach maximum extinction. The atomic cloud constitutes a cold thermal gas around the ONF. The atom that interacts significantly with the nanofiber mode does so for approximately 1.5 $\mu$s (see atomic transit measurements in  \cite{PhysRevA.92.013850}). Because the atomic cloud expansion reduces the density of atoms, we limit the probing time to 1.7 ms.  During this time we send a train of 200-ns probe pulses every 1.5 $\mu$s (approx. 1100 pulses). The probe beam is turned off and the MOT beams on. We reload the MOT for 20 ms and repeat the cycle. The average acquisition time for an experimental realization is around 5 hours, for a total of about $1 \times 10^{9}$ probe pulses.

When atoms are around the nanofiber, they tend to adhere to it due to van der Waals attraction. After a few seconds of exposing the ONF to rubidium atoms, it becomes coated with rubidium and light cannot propagate through. In order to prevent this, we use approximately $~$500 $\mu$W of 750 nm laser (Coherent Ti:Saph 899) during the MOT-on stage of the experimental cycle to create a repulsive potential that keeps the atoms away from the nanofiber surface. When the MOT beams turn off so does the blue beam, allowing the probed atoms to get closer to the ONF. We have also seen that 500 $\mu$W of blue detuned beam is intense enough to heat the nanofiber and accelerate atomic desorption from the surface.

Regarding the reduction of systematic errors, all the components of the magnetic field at the position of the MOT are carefully minimized. Using three sets of Helmoltz coils we reduce all residual field components to the level of 10 mG. This reduces low frequency quantum beats among different Zeeman sub-levels (with different $m_F$) that will shorten the apparent lifetime, and effects of atomic precession during the decay \textit{i.e.} the Hanle effect. The intensity of the probe pulse is kept much lower than the saturation intensity, in order to reduce the atomic excitation when the pulse is nominally off. Another systematic error is the lengthening of the measured lifetime due to radiation trapping, which is the multiple scattering of a photon between different atoms \cite{RT}. Light trapped in the sample can re-excite atoms near the ONF, creating the appearance of a longer atomic lifetime. We confirm that the atomic density is low enough by measuring the lifetime of atoms emitting into free space as a control measurment, similar to the approach followed in \cite{PhysRevA.57.2448}. The photons collected from emission into free space come mainly from atoms far away from the ONF surface, so their time distribution should give us the well known atomic lifetime $\tau _0=26.24(4)$ ns \cite{lifetime} in the absence of significant systematic error.  We also consider the modification of the probe polarization after being scattered by the nanofiber. However, given the symmetry of the problem, a horizontally polarized incoming beam does not change polarization after interacting with the ONF. On the other hand, vertically polarized light changes polarization in the transversal plane of the nanofiber. This leads to a different arrangement of dipoles aligned along $r$ and $\phi$ compare to a probe beam propagating unaltered, but does not change the overall distribution of dipoles aligned along both directions.

%%%%%%%%%%%%%%%%%%%%%%%%%%%%%%%%%%%%%%%%%%%%%%%%%%%%%

\section{Lifetime measurements} \label{data}

\begin{figure}
\includegraphics[width=0.45\textwidth]{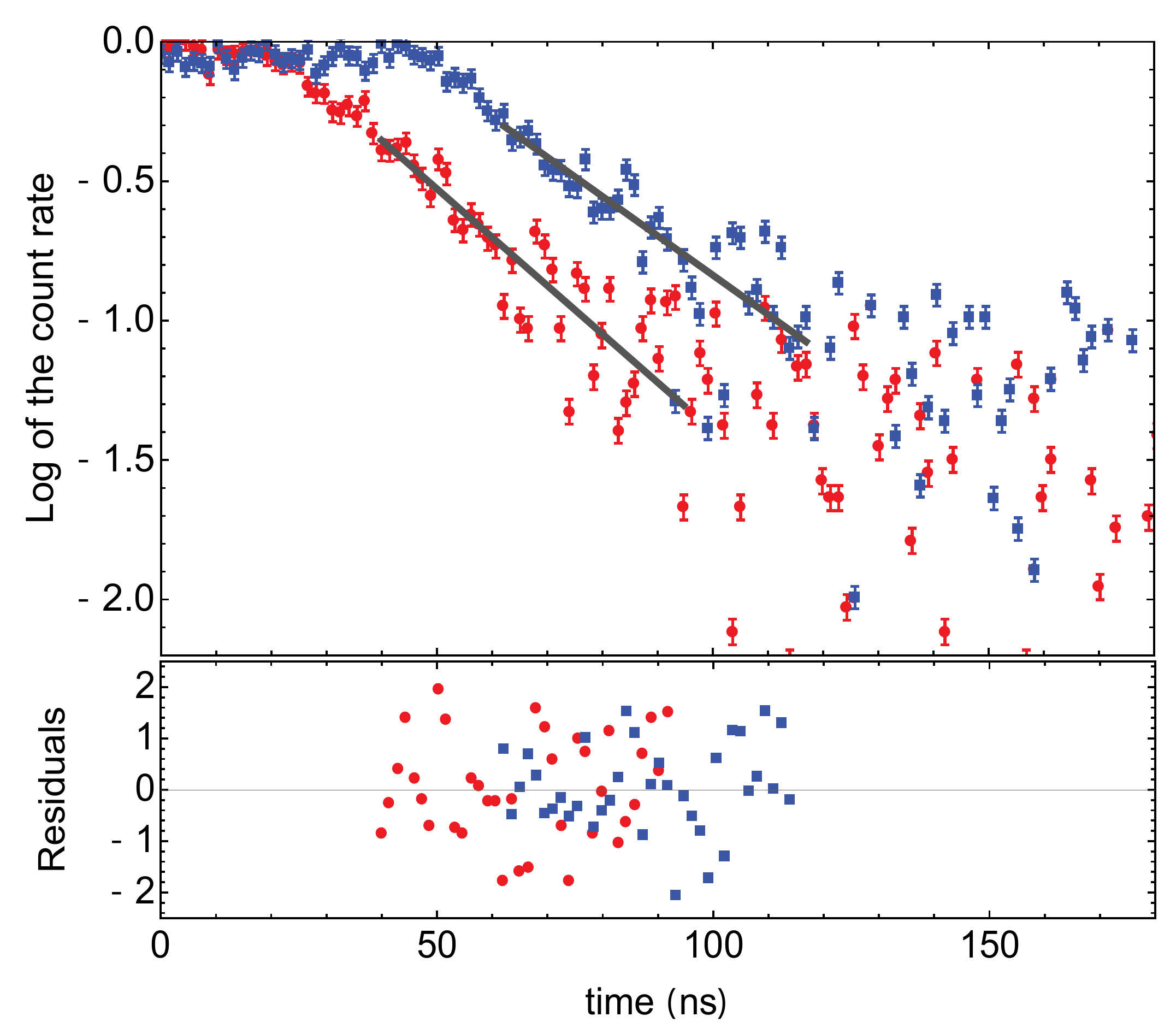}
\caption{Normalized time distribution of the collected photon count rate in logarithmic scale with a time bin of 1 ns. The red circles (blue squares) correspond to the data for vertically (horizontally) polarized probe light. The black solid lines are fits to exponential decays, and their residuals normalized to the standard deviation are displayed below the plot. The curves have been displaced 30 ns apart for clarity. }
\label{Decay Rates}
\end{figure}

We show the normalized time distribution of photons collected through the ONF mode in Fig. \ref{Decay Rates}. The red circles correspond to the data obtained for vertically polarized probe light, and the blue squares, for the horizontal one. The curves are horizontally shifted apart for clarity. The bin size is 1 ns and we typically have a thousand counts per bin at the peak. The error bars come from the statistical error of the data collection. The solid black lines are the fits to an exponential decay, and the plot underneath shows the corresponding normalized residuals. The fitting function is $A e^{-\gamma t}+O$, where the amplitude $A$ and the decay rate $\gamma$ are the only fitting parameters, and the offset parameter $O$ comes from the average value of the background at long times.

We vary the starting and ending point of the fitting curves and verify that as long as we are in a region within one to three natural lifetimes after the pulse turns off, there is no significant dependence on the chosen data points. Varying the end points did not change the obtained decay rate by more than 0.1\% . We consider only the fits with reduced $\chi^2$ between 0.9 and 1.5. The averaged decay rates extracted from these fits are $\langle\gamma\rangle_{v}/\gamma_{0}=1.088 \pm 0.015$ and $\langle\gamma\rangle_{h}/\gamma_{0}=0.943 \pm 0.014$ for the atoms driven by vertical and horizontal polarized probe light respectively. For these two data sets, the average of the measured free-space decay rates is $\langle\gamma_{0}\rangle/\gamma_{0}=0.989 \pm 0.012$, corresponding to atoms far away fro the ONF. The uncertainties represent the amount that the fitting parameter $\gamma$ has to be varied to changes the $\chi^2$ by plus or minus one. 

To study the systematic errors, we vary the magnetic field around 60 mG without observing a significant change on the decay rate. We also change the atomic density, and effects of radiation trapping bigger than the statistical errors appear when the density increases by a factor of three. The polarization of the probe pulse might also contribute with an error from a possible tilt of the ONF. We estimate this uncertainty to smaller than 10 mrad, and its effect into the total decay rate to be smaller than 0.1\%. 

Even though our signal to background should be in principle limited by the extinction ratio of the probe pulse (better than 1:170 after 20 ns), the signal is small enough that dark counts from the APDs become important and are our ultimate limiting factor. In our case the dark counts are around 500 counts per second, a factor of two higher than the specifications. However, the obtained signal is enough to measure a difference in the modified spontaneous emission decay rate for the two probe polarizations of almost 10 standard deviations.

%%%%%%%%%%%%%%%%%%%%%%%%%%%%%%%%%%%%%%%%%%%%%%%%%%%%%
\begin{figure*}
\includegraphics[width=1\textwidth]{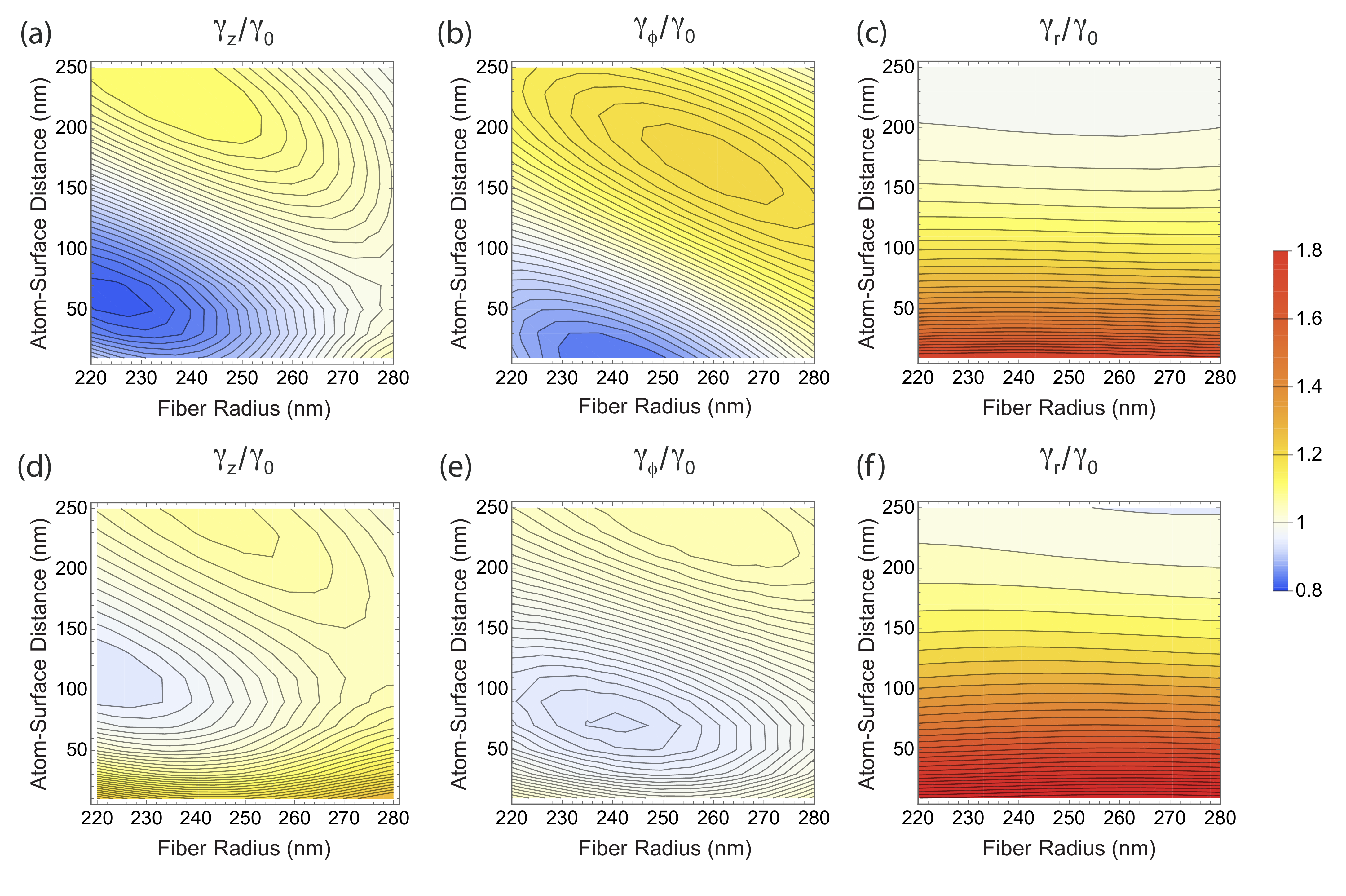}
\caption{Modification of the atomic spontaneous emission rate due to the presence of the ONF normalized by the free space decay rate. The results are displayed as function of the distance between the atom and the fiber surface, and the ONF radius. The three possible atomic dipole orientations can be along $z$, $\phi$, and $r$. (a)-(c) show the result of FDTD calculation. (d)-(e) Show the result of a mode expansion calculation}
\label{sims}
\end{figure*}

\section{Numerical Simulations for a Two-Level Atom}\label{NS}

Most of the literature about modified spontaneous emission rates considers two-level atoms, \textit{i.e.} classical dipoles, and we will follow that in this section.  We will discuss the ramifications of our multi-level atoms in a later section.

The radiative decay rate of an atom can be modified by the boundary conditions of the electromagnetic vacuum. We consider calculating this modification by two different approaches. Each is presumably equally valid and provides a different perspective and intuition of the problem \cite{hinds_cavity_1990}. In the first one, when the mode expansion of the electromagnetic field of the full space is known, the contribution of each mode to the spontaneous emission rate can be calculated using Fermi's golden rule (see Eq.(\ref{FGR})). In particular, the mode expansion of the vacuum electromagnetic field for an ONF has an analytical expression \cite{le_kien_spontaneous_2005}. A second strategy, useful when the modes are unknown or too complicated to compute analytically, is to solve the problem from classical electrodynamics. We calculate the modification of the radiated power of a classical dipole under equivalent boundary conditions, and take that to be the modification of the radiative decay rate of the atom \cite{Nano-Optics}
\begin{equation}
\frac{\gamma}{\gamma_0}=\frac{P}{P_0},
\end{equation}
where $\gamma$ and $\gamma_0$ are the modified and unmodified atomic decay rates respectively, and $P$ and $P_0$ are the classically calculated modified and unmodified radiative power. The modification of the radiative spontaneous emission is explained by the effect of the electric field reflected from the boundaries to the dipole position.

The latter approach allows us to develop an intuitive picture based on the idea that a two-level atom radiates as a linear dipole oscillating along the direction of the excitation field: When the atomic dipole is aligned along $z$ and $\phi$, parallel to the ONF surface, the radiated light can be reflected from the front and back interfaces created by the dielectric. These multiple reflections add at the position of the dipole affecting its emission. Because there is interference between reflections, the dipole radiation is sensitive to changes in the ONF radius. For these cases, the effect of the nanofiber can lead to enhancement or inhibition of the radiative spontaneous decay rate. On the other hand, for a dipole aligned along $r$ we can expect little radiation reflected from the back surface of the ONF to the dipole, given the radiation pattern. For this case, the decay rate depends mainly on the distance between the atom and the ONF surface and only slightly on the nanofiber radius. An alternative viewpoint is to consider image charges. The atomic dipole induces an image dipole inside the ONF aligned in the same axis and in phase. They radiate more power than the normal atomic dipole, producing an enhancement of the decay rate for the distances we consider.

Using the second strategy, based on a classical dipole, we calculate the modification of the atomic decay rate near an optical nanofiber as a function of the ONF radius and the distance from the atom to the nanofiber surface for different atomic dipole orientations. The calculation is performed numerically with a finite-difference time-domain (FDTD) algorithm \cite{taflove2005}. It considers the wavelength of the emitted light, and the nanofiber index of refraction to be $\lambda=780.241$ nm and  $n=1.45367$, respectively. The result of these calculations are shown in Fig. \ref{sims} (a)-(c). It shows the modification of the atomic decay rate as a function of distance to the nanofiber and radius of the nanofiber for dipoles aligned (in cylindrical coordinates) in the (a) $z$-direction, (b) $\phi$-direction, and (c)  $r$-direction relative to the ONF (as sketched in Fig. (\ref{sketch}) (b) and (c)). We identify these rates as $\gamma_{z}$, $\gamma_{\phi}$ and $\gamma_{r}$ respectively. The values of $\gamma/\gamma_0$ for this three cases are normalized so they are equal to one at large atom-surface distance.

The atomic decay rate of an atomic dipole aligned along $z$, Fig. \ref{sims} (a), is mostly inhibited close to the ONF surface compared to the free space decay rate, and it is highly dependent on the nanofiber radius. This is also true for dipoles aligned along $\phi$, Fig. \ref{sims} (b). For a dipole aligned along $r$, Fig. \ref{sims} (c), the decay rate is enhanced and depends mostly on the distance between the atom and the ONF surface and not on the nanofiber radius.

These results are compared with the calculations of the radiative lifetime using the electromagnetic field mode expansion (taken from Ref. \cite{le_kien_spontaneous_2005}) in Fig. \ref{sims} (d)-(f). We are interested in the limit where only the fundamental mode of the ONF can propagate, which is valid when the ONF radius is smaller than 284 nm for a wavelength of 780 nm. We observe that both calculations are qualitatively similar, but quantitatively different. The main discrepancy occurs at the fiber surface, where the mode expansion calculation seems to give a larger enhancement of the decay rate. The reason for the disagreement between both results is not understood. We have verified that other calculations based on finding the electric field at the position of the atom are in agreement with the mode expansion approach (compare Ref. \cite{le_kien_spontaneous_2005} and \cite{klimov_spontaneous_2004}). On the other hand, our FDTD calculations are in agreement with previous results using the same method \cite{verhart_single_2014}.

\section{Theoretical model} \label{TM}

The modification of the atomic spontaneous emission is a function of the position of the atom. Because the atoms are not trapped at a particular position, the measured decay time is a spatial average of the atomic distribution around the ONF. The main factor that determines such distribution is the van der Waals interaction between the atoms and the ONF. Moreover, atoms emit into the ONF mode with different probabilities, depending on their relative orientation and proximity to the nanofiber, altering the average of the decay time. We describe the necessary physical considerations to model the spatial average of the atomic distribution and the dipole orientation average corresponding to a multilevel atom.

\subsection{Van der Waals potential}

At short distances from the ONF the atoms feel an attractive force due to the van der Waals and Casimir-Polder (vdW-CP) potentials. These two potentials can be smoothly connected in a simple equation written as \cite{PhysRevA.77.042903,0953-4075-42-18-185006}
\begin{equation}
U_{g,e}(r)=-\frac{C^{(g,e)}_4}{r^3\left(r+C^{(g,e)}_4/C^{(g,e)}_3\right)},
\label{U}
\end{equation}
valid for the atomic ground (g) and excited (e) states, where $C_3$ and $C_4$ are the van der Waals and Casimir Polder coefficients of the atom interacting with the nanofiber. Using the procedure described in Ref. \cite{thesisJeff} we can obtain the value of these coefficients. For a $ ^{87}$Rb atom in the $5S_{1/2}$ ground level in front of an infinite half space fused silica medium, with index of refraction $n=1.45$,  the van der Waals and  Casimir-Polder coefficients are  $C^{(g)}_3=4.94 \times10^{-49}$ J$\cdot$m$^3$ and $C^{(g)}_4=4.47 \times10^{-56}$ J$\cdot$m$^4$ respectively. For the $5P_{3/2}$ excited state $C_3^{(e)}=7.05 \times10^{-49}$ J$\cdot$m$^3$ and  $C^{(e)}_4=12.2 \times10^{-56}$ J$\cdot$m$^4$. The vdW-CP potential affects the experimental measurement in two different ways, by reducing the local density of atoms and by shifting the atomic levels.

\begin{figure}
\includegraphics[width=0.45\textwidth]{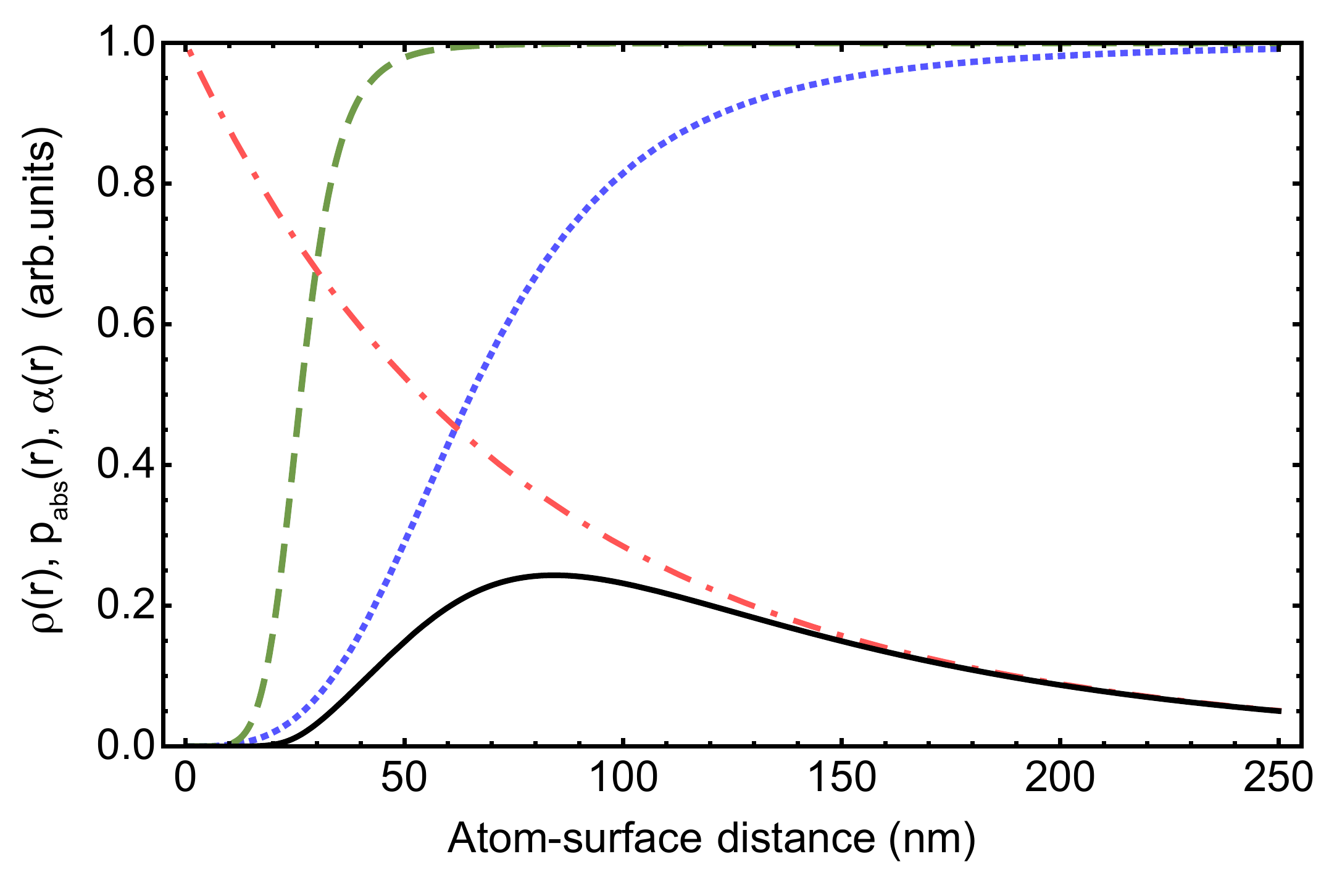}
\caption{Plot of the spatial dependence of $\rho(r)$ (dotted blue), $p_{\text{abs}}(r)$ (dashed green), and $\alpha(r)$ (dotted dashed red) in arbitrary units as a function of the atom-surface distance. The black solid line is the direct multiplication of these functions and it represents the distribution over which the spatial average is taken in a realistic experiment, as stated in Eq. (\ref{average-2level}).}
\label{distributions}
\end{figure}

By sending probe pulses to the entire atomic cloud, we actually measure a spatial average over an ensemble of atoms with a density distribution $\rho(r)$ at a radius $r$ from the ONF surface. The vdW-CP attraction accelerates the atoms reducing the local density  around the nanofiber, all of them initially in the ground state. Assuming only the radial degree of freedom and thermal equilibrium, a simple steady state density distribution can be obtained from the ideal gas law and energy conservation \cite{thesisJeff}, as  
\begin{equation}
\rho(r)\approx \rho_0 \frac{1}{1-U_g(r)/E},
\label{rho}
\end{equation}
where $\rho_0$ and $E=\frac{3}{2}k_B T$ are the atomic density and the average (kinetic) energy of the atoms far away from the fiber, with atoms typically at $T\approx$ 150 $\mu$K for our atomic cloud. By only considering $U_g$, we neglect the small fraction of atoms in the excited state. Fig. \ref{distributions} shows an example of this distribution (blue dotted line). This approximation agrees with previous analytical results \cite{PhysRevLett.99.163602}, and differs at most by 30\% with Monte Carlo simulations of atomic trajectories \cite{PhysRevA.77.042903}. 

The vdW-CP potentials also shift the atomic energy levels, affecting the probability to absorb the otherwise resonant probe beam as \cite{foot}
 \begin{equation}
p_{\text{abs}}(r)= \frac{N}{1+s+4(\frac{\delta(r)}{\gamma_0})^2},
\label{pabs}
\end{equation}
where $N$ is just a probability normalization factor and $\delta(r)=(U_e(r)-U_g(r))/2\pi\hbar$ is the detuning induced by the ONF,  which for us is always red shifted. This distribution is plotted with a green dashed line in Fig. \ref{distributions}, neglecting $s$ since we work in the low saturation limit.

\subsection{Coupling into the waveguide}

Another factor to consider when measuring the spontaneously emitted light into the ONF, is the fact that atoms that are closer to the nanofiber surface contribute more to the measured signal than those further away. This effect is characterized by the emission enhancement parameter 
 \begin{equation}
 \alpha(r)=\frac{\gamma_{\text{wg}}(r)}{\gamma_0}.
 \label{alpha}
 \end{equation}
This factor is different from the more commonly used coupling efficiency $\beta(r)=\gamma_{\text{wg}}(r)/\gamma(r)$ \cite{PhysRevA.80.011810}. $\alpha(r)$ is proportional to the total number of photons emitted into the guided mode, and $\beta(r)$ is the percentage of photons emitted into the mode relative to the total number of emitted photons. The difference between $\alpha$ and $\beta$ becomes clear with the following example: When the coupling efficiency $\beta(r)$ is very large, close to one, most of the emitted photons couple to the waveguide. However, the total number of photons emitted into the waveguide can still be close to zero if the total spontaneous emission where to be greatly inhibited, $\gamma \ll \gamma_0$ (which is not our particular case). The amplitude of the signal measured through the guided mode is then represented by the emission enhancement parameter $\alpha(r)$.

An analytical expression for $\gamma_{\text{wg}}(r)$ can be found in the literature \cite{le_kien_anisotropy_2014, le_kien_spontaneous_2005}, and it is proportional to the norm squared of the evanescent electric field, as expressed in Eq. (\ref{FGR}). For a single mode ONF, the spatial dependence of each component of the evanescent electric field is given by the sum of one or two modified Bessel function of the second kind $K_{i}(qr')$ of order $i=0,1,2$; where $r'=r_0+r$, and $r_0$ and $r$ are the ONF radius and the radial position from the ONF surface;  $q=\sqrt{\beta^2-k^2}$ is the transverse component of the wave vector, $\beta$ is the field propagation constant in the ONF, and $k=2 \pi/\lambda$ is the free space wavenumber. For our particular case of an ONF with radius $r_0=230$ nm propagating a field of wavelength $\lambda=780$ nm, $q=0.56k$. Provided that $qr'>1$ away from the ONF surface, we can simplify the calculation with the asymptotic expansion of $K_i(qr')\approx\sqrt{\pi/2qr'}e^{-qr'} $, and approximate the spatial dependence of $\gamma_{\text{wg}}(r)$ as \cite{PhysRevA.92.013850}
\begin{equation}
\alpha(r)=\frac{\gamma_{\text{wg}}(r)}{\gamma_0}\propto \frac{1}{r_0+r} e^{-2(0.56 k r)},
\end{equation}
approximation that has been tested against exact numerical results with excellent agreement. Any other constant pre-factor will not contribute to the final average after the appropriate normalization. Its spatial distribution is plotted as the red dotted and dashed line in Fig. \ref{distributions}. For our experimental parameters $\alpha\approx0.2$ at the ONF surface.

Atomic dipoles aligned along different direction will couple to the guided mode with different strengths. We denote the emission enhancement parameter with a subindex to specify the alignment of the emitting dipole as $\alpha_{i}(r)$ with $i \in \{z,\phi,r\}$. It can be shown, from the calculation in Ref. \cite{le_kien_spontaneous_2005}, that to a good approximation $\alpha_{z}\approx\alpha_{\phi}\approx\alpha_{r}/3$ for our ONF, independent on the radial position of the atom. The different coupling strength for atomic dipoles aligned along $r$ comes from the fact that the radial component of the guided field is discontinuous and larger than the others due to the dielectric boundary conditions. 

Figure \ref{distributions} shows the spatial dependence of each one of the described distributions that affect the measured average decay rate. The black solid line in the plot represents the direct multiplication of them. This effectively describes the probability of observing a photon emitted from an atom at a position $r$ into the ONF guided mode. Noticed that for a given ONF the only experimentally tunable parameter for the final distribution is the temperature of the atomic cloud. The atomic distribution is weakly dependent on the temperature in Eq.(\ref{rho}).

\subsection{Averaged signal}

The measured signal is an average of atoms decaying at different rates. If these decay rates are close enough to each other the measured decay rate is approximately equal to the spatially averaged decay rate $\langle\gamma\rangle$. As a proof let us consider that the decay rates differ by a small quantity $\epsilon$ with a distribution $g(\epsilon)$, then the measured signal is given by
\begin{equation}
\int d\epsilon g(\epsilon)e^{-\gamma(1+\epsilon)t}=e^{-\gamma t} \int d\epsilon g(\epsilon) e^{-\gamma \epsilon t}.
\end{equation}
For small $\epsilon$ (and short times) the exponential of order $\epsilon$ can be expanded in series, averaged, and exponentiate again, to obtain
\begin{equation}
e^{-\gamma(1+\langle \epsilon\rangle)t}=e^{-\langle \gamma\rangle t}
\end{equation}

The measured decay rate $\langle \gamma\rangle$ is a spatial average of the actual decay rate weighted by the atomic density distribution $\rho(r)$, the excitation probability $p_{\text{abs}}(r)$ and the emission enhancement parameter $\alpha(r)$.
\begin{equation}
\langle\gamma\rangle=\frac{\int \gamma(r)\alpha(r) \rho(r) p_{\text{abs}}(r)r dr}{\int \alpha(r) \rho(r) p_{\text{abs}}(r)r dr}.
\label{average-2level}
\end{equation}

In the particular case of driving the atoms with light polarized vertically, there are atomic dipoles oriented along $r$ and $\phi$ (see Fig. \ref{sketch} (c)). In this case the proper $\alpha_i$ has to be taken into account to obtain the averaged signal.

\section{Consideration of the multi-level atomic structure} \label{multilevel}

The radiation pattern of a real atom differs from that of an ideal linear dipole. A multi-level atom (with more than one Zeeman sub-level $m_F$ in the ground state) can decay to a ground state through a $\pi$- or $\sigma$-transition, when $\Delta m_F=0$ or $\Delta m_F=\pm 1$ respectively.  In our case, we consider the quantization axis to be along the direction of the linear polarization of the probe, and $\pi$ and $\sigma$ are with respect to this quantization axis.

We model the decay rate of a multi-level atom by an incoherent superposition of linear dipoles \cite{Polder1976} that describes the real radiation pattern.  An atom decaying through a $\pi$-transition is described by a linear dipole aligned along the probe polarization axis. An atom decaying through a $\sigma$-transition, is considered as a linear dipole rotating in the plane perpendicular to the probe polarization axis. The atomic decay rate depends on the norm squared of the dot product of the electric field and the dipole polarization (see Eq. (\ref{FGR})). This implies that the decay rate of a rotating dipole ($\sigma$-transition) is the incoherent sum of the decay rate of two orthogonal linear dipoles oscillating in the rotation plane.

All the information necessary to calculate the decay rate of different transitions of a real atom are then calculated from the decay rates of classical linear dipoles, in the spirit of Fig. \ref{sims}. To  calculate the total decay rate we need to know the branching ratios of the transitions. We denote the probability of decay through a $\pi$- or $\sigma$-transition as $P_{\pi}$ and $P_{\sigma}$ respectively. This depends on the state preparation of the atoms.

The measured signal will be a spatial average over the contribution of such classical dipoles, weighted according to the coupling efficiency of each dipole into the waveguide. Considering this, the spatially averaged decay rate can be obtained using Eqs. (\ref{rho}), (\ref{pabs}) and (\ref{alpha}), as
\begin{equation}
\langle\gamma\rangle_{\pi}=\frac{\int \left(\gamma_{\pi}(r)P_{\pi}\alpha_{\pi}(r) +\gamma_{\sigma}(r)P_{\sigma}\alpha_{\sigma}(r)\right) \rho(r) p_{\text{abs}}(r)r dr}{\int \left(P_{\pi}\alpha_{\pi}(r) +P_{\sigma}\alpha_{\sigma}(r)\right) \rho(r) p_{\text{abs}}(r)r dr}.
\label{average-multilevel}
\end{equation}
where $\gamma_{i}(r)$ with $i \in \{\pi,\sigma\}$ are obtained from the numerical simulation displayed in Fig. \ref{sims}. The subscript $\pi$ in the spatial average denotes the polarization of the probe beam that drives the atomic transition.

\section{Comparison between theory and experiment} \label{comparison}

We can compare the measurements with the theoretical simulations calculating the average $\langle\gamma\rangle/\gamma_0$ by introducing the numerical values of $\gamma(r)/\gamma_0$, displayed in Fig \ref{sims}, into Eq. (\ref{average-2level}) for a particular ONF radius. It is important to notice that when we realize the experiment probing the atoms with horizontally polarized light we are measuring the spatially averaged decay rate for an atomic dipole aligned along $z$. For atoms driven by a vertically polarized probe pulse we measure a decay rate that is averaged over the different dipole alignments in addition to the spatial average (along $\phi$ and $r$ as is is shown in Fig. \ref{sketch} (c)). This means that we can not separately measure the decay rate for dipoles aligned along $r$ and $\phi$.

\begin{figure}
\includegraphics[width=0.45\textwidth]{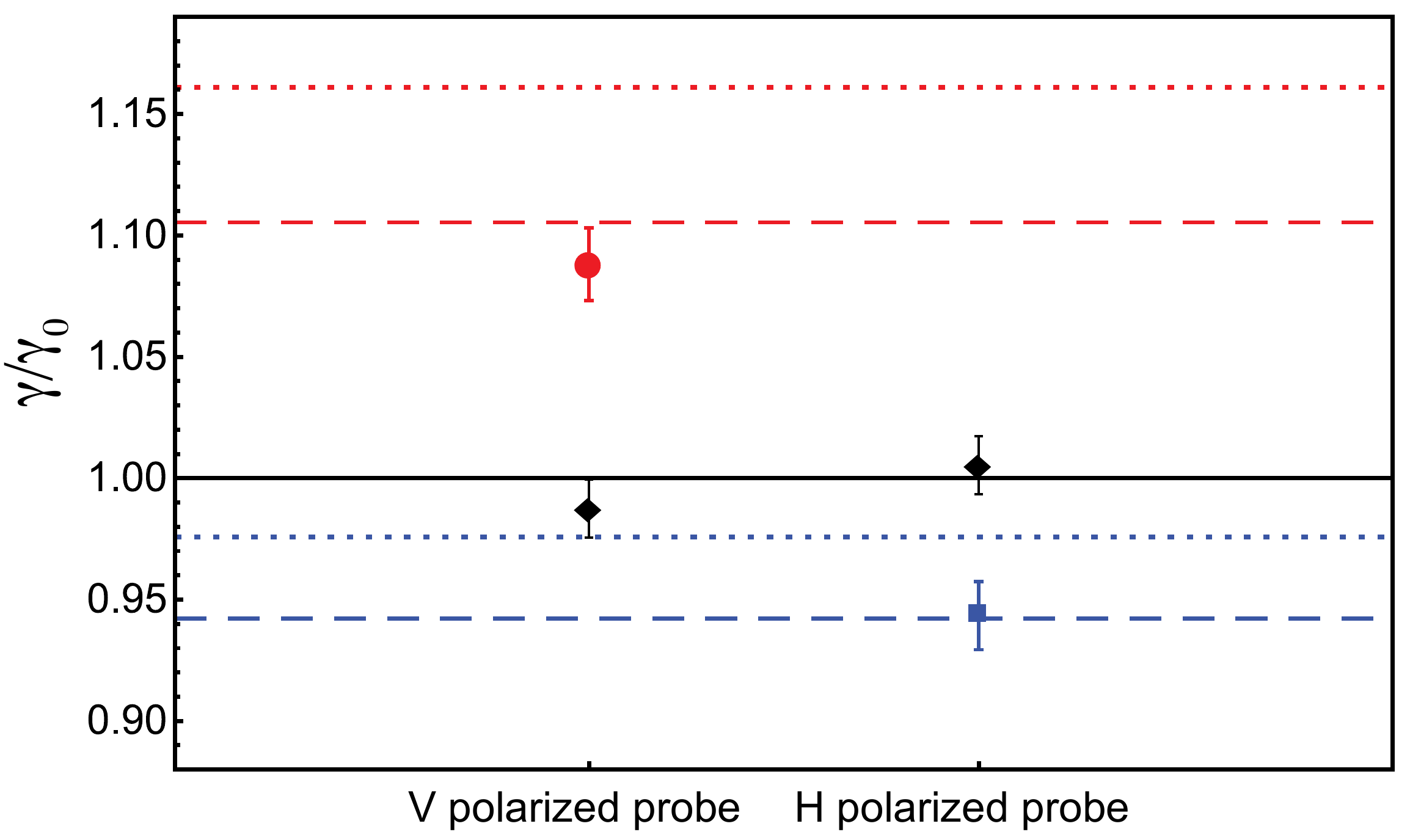}
\caption{Normalized decay rates for different polarizations of the probe with respect to the nanofiber. The red circle (blue square) corresponds to the measured modified lifetime of atoms driven by vertically (horizontally) polarized probe light. The black diamonds are the simultaneously measured free space decay time for each configuration.The solid black line is the expected decay rate in free space.  The dashed blue and red lines are the calculated values from the two-level atom FDTD calculation for a horizontal polarized probe and a vertically polarized probe respectively. The dotted lines are the calculated values from the two-level atom mode expansion calculation. Both calculation are done considering the spatial average in Eq. (\ref{average-2level}).}
\label{results}
\end{figure}

Figure \ref{results} shows a comparison between the measurements and the numerical simulations for a two-level atom. It shows the extracted atomic decay rates for both experimental configurations normalized by the free space one. The blue lines are the calculated value of the modified decay rate $\langle\gamma\rangle_{h}/\gamma_0$, corresponding to the probe beam horizontally polarized, to be compared with the experimental value (blue square). The blue dashed line corresponds to the FDTD calculation and the dotted one the mode expansion calculation. The red lines are the calculated values of the modified decay rate $\langle\gamma\rangle_{v}/\gamma_0$, corresponding to the probe beam vertically polarized, to be compared with the experimental value (red circle). The red dashed line corresponds to the FDTD calculation and the dotted one the mode expansion calculation. For each experimental realization we simultaneously measure the modified atomic decay rate of atoms close to the ONF and the free space decay rate from atoms in the MOT, where the great majority of them are away from the nanofiber. The black diamonds in Fig. \ref{results} are the measurements of the decay time into free space for the two different polarizations. When the measured decay rate into free space is off by more than few percent of the expected value, because for example an unexpected fluctuation of the atom density, the data collected through the ONF mode was discarded.

For a multilevel atom we have to consider its initial state. During the period the probe beam is on, the atoms get pumped into a particular ground state. The steady state solution for optical pumping the $F=2\rightarrow F'=3$ transition of $^{87}$Rb with linearly polarized light is biased towards the $m_F=0$ state (the fractional populations are approx. 0.04, 0.24, and 0.43 for $|m_F|$ equal 2, 1, and 0 respectively). A $\pi$ excitation (linearly polarized) of such initial state, will lead to probability $P_{\pi}=0.55$ of emitting $\pi$ radiation and $P_{\sigma}=0.45$ of emitting as a $\sigma$ radiation (circularly polarized). This effect has to be taken into account when calculating the averaged decay rate using Eq. (\ref{average-multilevel}). Considering this, neither of our calculations, the FDTD nor the mode expansion, predict the measured values. They tend to be approximately equal to the natural decay rate for both polarizations. Even if the population distribution is different from the calculated optical pumping values, almost all population distributions tend towards producing more isotropic distributions than a linear dipole, as any amount of sigma polarization reduces the angular contrast.

The fact that the radiation pattern of a real atom is more isotropic than the one of a linear dipole questions the idea of having any significant alignment dependent effect. The spontaneous decay of an atom to a particular ground states does not depend on the atomic dipole coherence induced by the excitation, but it depends only on the branching ratio of the decay of the excited state (proportional to the square of the Clebsch--Gordan coefficients). This is an outstanding puzzle: our measurements are in good agreement with a radiating linear dipole and in disagreement with the radiation expected for the actual multi-level atom.  We note that the problem of the radiation of multi-level atoms with degenerate ground states near systems with a modified environment has not been thoroughly addressed or experimentally studied.

%%%%%%%%%%%%%%%%%%%%%%%%%%%%%%%%%%%%%%%%%%%%%%%%%%%%%

\section{Discussion} \label{discussion}

The numerical calculations have only the ONF radius and the atomic cloud temperature as adjustable parameters to match the experimental results, the former one showing a stronger effect in the expected decay rate. The fact that the decay time of an atomic dipole oriented along the fiber is strongly dependent on the nanofiber radius, gives us a possible method to measure this radius. From the error bars in the collected data,  and considering the dependence on the nanofiber radius of the simulations, we determine the radius to be $235\pm 5$ nm, based on the two-level atom FDTD calculation. This value is close the estimated ONF radius from the fabrication (240 nm $\pm$ 20) . Having fixed the nanofiber radius, there are no free parameters in the calculation of $\langle\gamma\rangle_{v}/\gamma_0$ (red dotted line in Fig. \ref{results}) other than the MOT temperature, set to be 150 $\mu$K. 

When we calculate the spatial average, as is explained in Sec.\ref{TM}, we make a series of approximations. These include the van der Waals and Casimir-Polder coefficients calculated for a dielectric plane instead of a cylinder in Eq.(\ref{U}), an equilibrium distribution of the atomic density in Eq.(\ref{rho}), and the asymptotic expansion of the guided mode for determining  the shape of $\alpha(r)$ in Eq.(\ref{alpha}). However, when we vary all these quantities ($C_3$, $C_4$, $\rho$, and $\alpha$) for 20\% of their values, the final averaged decay rate does not change by more than the estimated error bars of the measurements. 

We can use the physical model presented in this paper to design other experiments. From Fig. \ref{sims} (a)-(c) we can see that by positioning the atoms, for example at 50 nm from a 230 nm radius ONF, we can create atomic states can go from an approximately  $40 \%$ enhancement of the spontaneous emission to $20\%$ inhibition. This is possible by using only the atomic dipole alignment as a tuning knob for its coupling to the mode of the nanofiber and the environment. We can also use ONFs that support higher order modes to allow us to have a better control of the probe polarization, using the evanescent field of guided light, which can be used to drive different atomic dipole orientations, such as purely radial or azimuthal.

The discrepancy between different calculations has yet to be understood. It is also necessary to develop better physical picture that explains the measured behavior for a real multi-level atom near an ONF. This problem bring fundamental questions that need to be revised and systematically study in the future, crucial for any future application of multi-level atoms coupled to optical waveguides.

%%%%%%%%%%%%%%%%%%%%%%%%%%%%%%%%%%%%%%%%%%%%%%%%%%%%%

\section{Conclusion}\label{conclusion}

We have experimentally observed the modification of the rate of spontaneous emission of atoms near an optical nanofiber and its dependence on the atomic dipole alignment. The experiment is implemented by placing an ONF at the center of a cold atomic cloud. A linearly polarized resonant probe pulse drives the atomic dipoles in a particular alignment. We measured the time distribution of spontaneously emitted photons into the ONF to obtain the atomic lifetime. The modification of the atomic lifetime is measured for different probe polarizations in order to show the dependence on the atomic dipole alignment of the spontaneous emission rate. 

A physical model of the experiment is also presented an used to perform a numerical calculation of the modification of the spontaneous emission rate. This shows a good quantitative agreement with the experimental measurements considering two-level atoms. Some basic physical aspect remain elusive, regarding the multi-level structure of a real atom. This work clearly demonstrates that there are open problems that need further investigation - a perhaps surprising conclusion given the fundamental nature of the simple problem of an atom radiating near a dielectric. A better understanding of the problem will allow us to extend this knowledge to more general cases. With this knowledge of how atomic properties change under different conditions, we can start implementing a new toolbox for precisely manipulating and controlling atoms coupled to optical waveguides.

%%%%%%%%%%%%%%%%%%%%%%%%%%%%%%%%%%%%%%%%%%%%%%%%%%%%%
\
\section*{Disclaimer}
Any mention of commercial products in a publication having NIST authors is for information only; it does not imply recommendation or endorsement by NIST.

\section*{Acknowledgments}
We would like to thank H. J. Charmichael and A. Asenjo-Garcia for their valuable contributions to the discussion. This research was supported by the National Science Foundation of the United States (NSF) (PHY-1307416);  NSF (CBET-1335857); NSF (PHY-1430094); and the USDOC, NIST, Joint Quantum Institute (70NANB16H168).

\bibliography{ref}

\end{document}